\documentclass{webofc}
\usepackage[varg]{txfonts}   % Web of Conferences font
\usepackage{color}

\usepackage{subfigure}
\woctitle{New Frontiers in Physics 2014}
\usepackage{lineno}
\begin{document}

\newcommand{\prepnum}[1]{\gdef\@prepnum{#1}}
\prepnum{\bf CLICdp-Draft-2014-019}
\title{CLICdp Overview}
%
% subtitle is optionnal
%
\subtitle{ Overview of physics potential at CLIC}

%\author{ Lucie Linssen\inst{1}\fnsep\thanks{\email{lucie.linssen@cern.ch}}
 %\and     
\author{Aharon Levy\inst{1}\fnsep\thanks{\email{levyaron@post.tau.ac.il}}\thanks{This work was partially supported by the EU AIDA project and by the German-Israel Foundation(GIF)} 
\\On behalf of the CLICdp Collaboration
%\and
 %       Third author\inst{3}\fnsep\thanks{\email{Mail address for last
  %          author if necessary}}
        % etc.
}

%\institute{CERN, Geneve, Switzerland
%\and
\institute       { Tel Aviv University, Tel Aviv, Israel   }
%\and
 %          Last address
         
\abstract{
 CLICdp, the CLIC detector and physics study, is an international collaboration presently composed of 23 institutions. The collaboration is addressing detector and physics issues for the future Compact Linear Collider (CLIC), a high-energy electron-positron accelerator which is one of the options for the next collider to be built at CERN. Precision physics under challenging beam and background conditions is the key theme for the CLIC detector studies. This leads to a number of cutting-edge R\&D activities within CLICdp. The talk includes a brief introduction to CLIC, accelerator and detectors, hardware R\&D as well as physics studies at CLIC. }

\maketitle

\section{Introduction}
\label{intro}

The Compact Linear Collider detector and physics (CLICdp) collaboration~\cite{CLICdp}  is a detector and physics R\&D international collaboration which includes 23 institutes from a large geographic area. 
%(see Table~\ref{tab:CLICdp} at the end of this writeup). 

An overview of the activities of CLICdp~\cite{LL-CDR} is presented in this talk. It first describes shortly the Compact Linear Collider (CLIC)~\cite{CLIC} with an overview of the physics scope which would drive the energy staging of the collider. This is followed by a description of the detector requirements and experimental conditions dictated  by the structure of the beam and by the physics program. The CLIC detector concept, the subdetectors and the respective R\&D work will be discussed. Next, a description is given  of the CLIC physics capabilities, including examples of benchmark studies based on full detector simulations with overlay of beam-induced background processes. The examples of physics capabilities cover the following physics topics: Higgs, Top and New Physics.

\section{The CLIC accelerator}

\begin{figure}[h!]
\centerline{\includegraphics[width=0.8\textwidth]{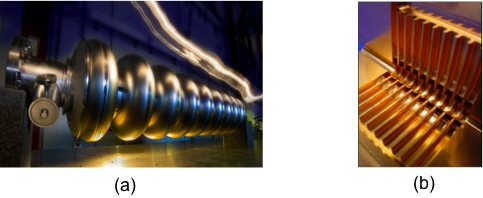}}
\caption{Superconducting RF cavities for the ILC (a) and  normal conducting acceleration structures for CLIC (b).}
\label{fig:ILC-CLIC}
\end{figure}

Currently there are two TeV-scale linear $e^+e^-$ colliders under development, CLIC~\cite{CLIC} and ILC~\cite{ILC}. Both machines are designed for high luminosities, of the order of 10$^{34}$~cm$^{-2}$sec$^{-1}$. The ILC uses superconducting RF cavities, 
 while CLIC uses a two-beam accelerator scheme at room temperature (Fig~\ref{fig:ILC-CLIC}).  A high current ($\sim$100~A) low-energy drive-beam serves as an RF source for the low-current ($\sim$1~A) high-energy main beam. 

\begin{figure}[h!]
\centerline{\includegraphics[width=0.8\textwidth]{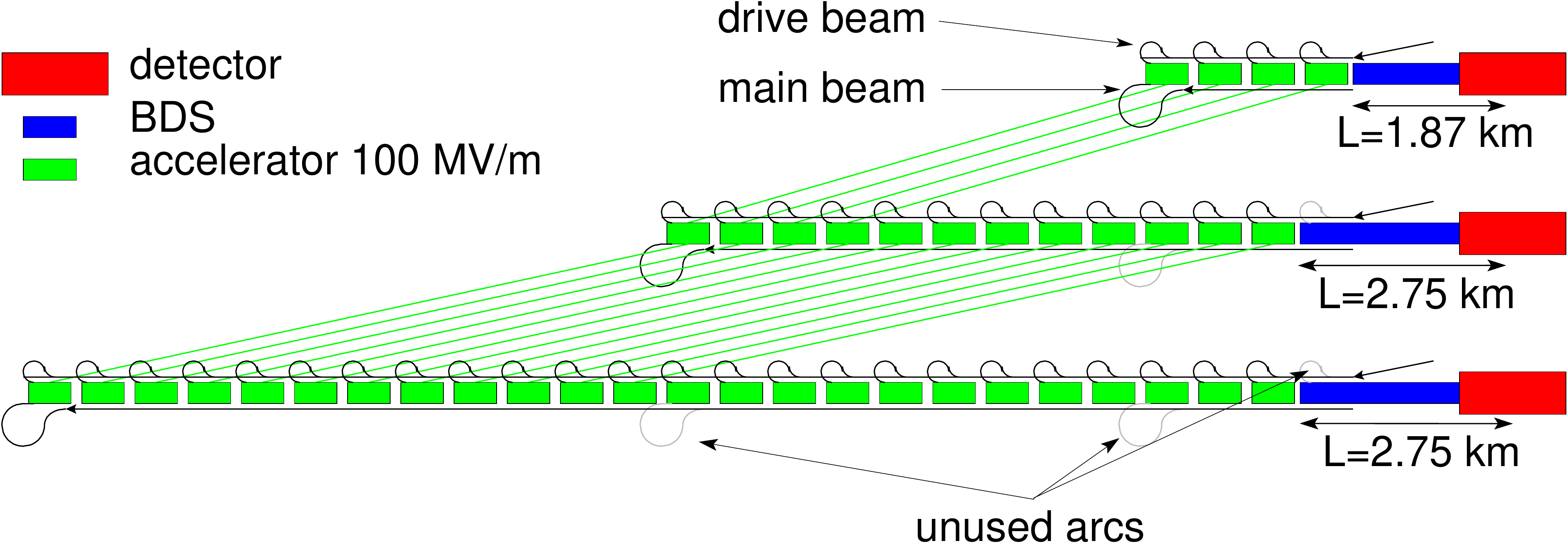}}
\caption{Simplified upgrade scheme for a CLIC staging scenario.}
\label{fig:staging}
\end{figure}

\begin{figure}[h!]
\centerline{\includegraphics[width=0.8\textwidth]{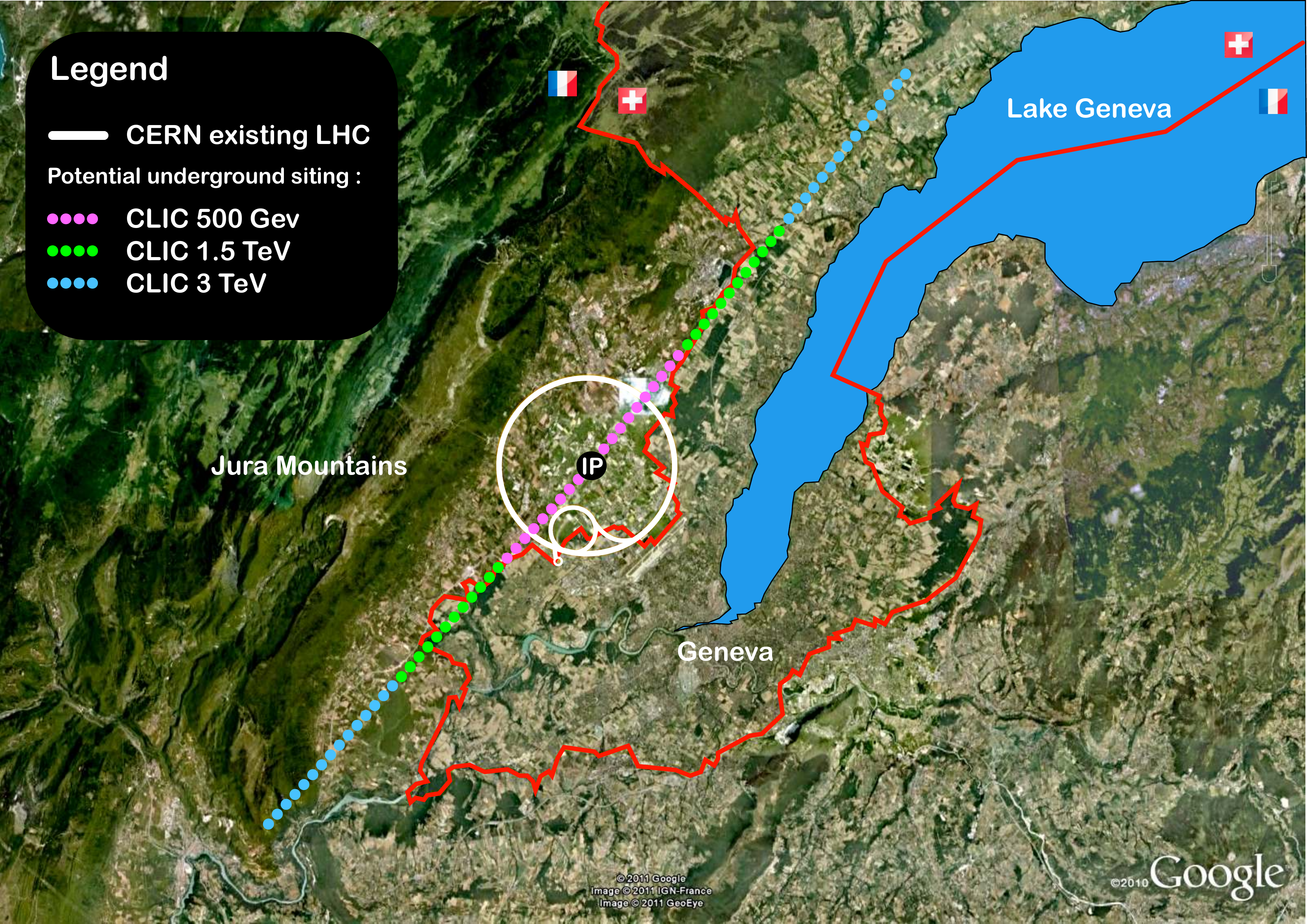}}
\caption{CLIC footprints near CERN, showing various implementation stages.}
\label{fig:CLIC-implementation}
\end{figure}

The ILC has a gradient of 32~MV/m
while CLIC has a gradient of 100MV/m and therefore can be a shorter machine for the same centre-of-mass (cms) energy. The ILC is designed for up to 500~GeV, upgradable to 1~TeV. CLIC is designed for up to 3~TeV and will be staged. The staging would be in three steps, for example 500~GeV, 1.4~TeV and 3~TeV. In one of the scenarios, based on accelerating units of 100~MeV/m, it will start as a 11.4~km long collider at 500~GeV, increase its energy to 1.5~TeV, resulting in a 27.2~km machine, and finally extend to a length of 48.3~km for a cms energy of 3~TeV (see Fig.~\ref{fig:staging}).
More details about the CLIC parameters for the three scenarios are listed in Table~\ref{tab:CLIC-stages}.
\begin{table}[tb]
\centering
\caption{Parameters for the CLIC energy stages.}
\label{tab:CLIC-stages}       % Give a unique label
% For LaTeX tables you can use
\begin{tabular}{|l|l|l|l|l|l|}
\hline
{ \color{red} {\bf Parameter}}& {\color{red}{\bf Symbol} }&  {\color{red}{\bf Unit} } &  {\color{red}{\bf Stage 1} } & {\color{red}{\bf Stage 2} }  &  {\color{red}{\bf Stage 3} } \\\hline \hline
Centre-of-mass energy& $\sqrt{s}$& GeV&  500& 1500 & 3000 \\ \hline
Repetition frequency & $f_{rep}$ & Hz & 50 & 50 & 50 \\ \hline
Number of bunches per train & $n_b$ &  & 312 & 312 & 312 \\ \hline
Bunch separation & $\Delta_t$& ns & 0.5 & 0.5 & 0.5 \\ \hline
Accelerating gradient & $G$ & MV/m &  100 & 100 & 100 \\ \hline
Total luminosity & $\cal{L}$ & 10$^{34}$~cm$^{-2}$s$^{-1}$ & 1.3 & 3.7 & 5.9 \\ \hline
Luminosity above 99\% of $\sqrt{s}$&$\cal{L}_{\rm 0.01} $&10$^{34}$~cm$^{-2}$s$^{-1}$&0.7&1.4&2 \\\hline
Main tunnel length &  & km & 11.4 & 27.2 & 48.2 \\ \hline
Charge per bunch & $N$ & 10$^9$ & 3.7 & 3.7 & 3.7 \\ \hline
Bunch length & $\sigma_z$ & $\mu$m & 44 & 44 & 44 \\ \hline
IP beam size & $\sigma_x/\sigma_y$ & nm & 110/2.6 &  $\approx$60/1.5 & $\approx$40/1 \\ \hline
Normalised emittance (end of linac) & $\epsilon_x/\epsilon_y$& nm&  - & 660/20 & 660/20 \\ \hline
Normalised emittance &  $ \epsilon_x/\epsilon_y$& nm& 660/25 & - & - \\ \hline
Estimated power consumption & $P_{wall}$ & MW& 235 & 364 & 589 \\ \hline
\end{tabular}
\end{table}
 CLIC is clearly an ambitious project which requires high power consumption but, as will be discussed later, it has the potential of producing exciting physics results. A possible implementation of such a machine near CERN is shown in Fig~\ref{fig:CLIC-implementation}.

The first stage will allow for guaranteed precision Higgs and top physics, so the energy of the first stage will be chosen near the $t\bar{t}$ threshold.  The total integrated luminosity of $\sim$500~fb$^{-1}$ required
at this energy will require to run for four years. Four more years will be needed to collect $\sim$1.5~ab$^{-1}$ of
 integrated luminosity at the 1.4~TeV stage, for more precision Higgs physics, and, hopefully following LHC discoveries, precision measurements of new physics phenomena Beyond the Standard Model (BSM) or indirect searches of BSM through high-precision observables. The 3~TeV stage will allow for a decent measurement of the elusive Higgs self-coupling, and for additional accurate BSM measurements and searches.

One possible time-line for the CLIC machine would be to have a (hopefully) positive decision in 2018-19, followed by a 4-5 year preparation phase and a start of construction in 2024-25 alongside with detectors construction and start of operation in 2032-33.

\section{CLIC experimental environment}

\subsection{The beam structure}

\begin{figure}[h!]
\centering
\includegraphics[width=0.8\textwidth]{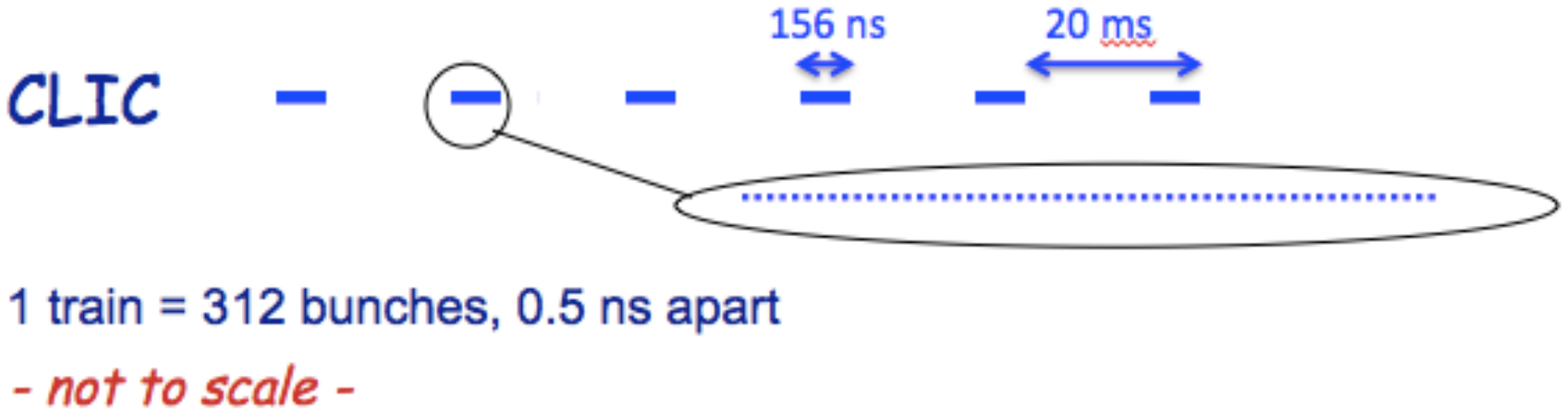}
\vspace*{-3cm}
\caption{The beam structure of CLIC. There are 312 bunches in a train, the bunches being 0.5~ns apart, the train length is 156~ns and the trains are separated by 20~ms.}
\label{fig:beam-structure}       
\end{figure}

The expected cross sections of the most interesting processes at the three stages are of the order of fb, so high luminosities are needed. In order to achieve such high luminosities at a linear collider, very small beams and a high beam repetition  rate are needed. It will have 50 bunch trains per second, each train will be 156~ns long, and will include 312 bunches each separated by 0.5~ns (see Fig~\ref{fig:beam-structure}). This will give a very low duty cycle which can be used to power pulse the detector components (see later). This reduces power dissipation and significantly reduces on-detector cooling infrastructure.
%\vspace{-3cm}
Beam sizes will be very small. With such a beam, detailed in Table~\ref{tab:beam-details}, we expect on average less than one physics event per bunch train.
\begin{table}[h!]
\centering
\caption{Some parameters of a 3~TeV CLIC beam. BX stands for bunch crossing.}
\label{tab:beam-details}       % Give a unique label
% For LaTeX tables you can use
\begin{tabular}{|l|c|}
\hline
 & CLIC at 3~TeV \\\hline
L(${\rm cm}^{-2}{\rm s}^{-1}$) & $5.9\times 10^{34}$ \\\hline
BX separation & 0.5~ns \\\hline
\#BX/train & 312 \\\hline
train duration & 156~ns \\\hline
repetition rate & 50~Hz \\\hline
duty cycle& 0.00078\% \\\hline
$\sigma_x/\sigma_y$ & 45/1~nm \\\hline
$\sigma_z$ &  44~$\mu$m\\\hline
\end{tabular}
% Or use
%\vspace*{5cm}  % with the correct table height
\end{table}
This means that no trigger is needed and all events can be recorded.

\subsection{Beam-induced background}

\begin{figure}[h!]
\begin{center}
\includegraphics[width=0.3\textwidth]{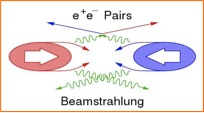}
\caption{Pair production through the Beamstrahlung process.}
\label{fig:beamstrahlung}
\end{center}
\end{figure}

The small bunch sizes required to achieve high luminosity at CLIC give rise to strong electromagnetic radiation, so-called beamstrahlung, from the electron and positron bunches in the high field of the opposite beam. This process is depicted in Fig.~\ref{fig:beamstrahlung} and gives rise to the creation of numerous incoherent electron-positron pairs, predominantly produced in very forward direction and with low transverse momenta.
 The large majority of particles from incoherent pairs leave the detector in the forward region, where some of them give rise to back-scattering off the most forward detector elements. A small fraction of the particles from incoherent pairs reaches the central detector region directly.

The average energy loss, $\delta E$, of a beam particle due to beamstrahlung is, in first approximation,
\begin{equation}
\delta E  \propto \frac {N^2 \gamma_L} {(\sigma_x+\sigma_y)^2\sigma_z} ,
\label{eq:beamstrahlung}
\end{equation}
where $N$ is the number of bunch particles, $\gamma_L$ is the particle Lorentz factor and $\sigma_x,\sigma_y,\sigma_z$ are the beam dimensions in $x,y$ and $z$ coordinates. This equation shows that the energy loss, and therefore the number of radiated photons, depends on the machine parameters and increases with decreasing beam dimensions.

\begin{figure}[h!]
\begin{center}
\includegraphics[width=0.3\textwidth]{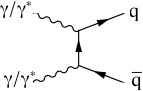}
\caption{Diagram describing the two-photon to hadrons process.}
\label{fig:gg-diagram}
\end{center}
\end{figure}

Another source of background is the $\gamma \gamma \to$ hadrons process, through the diagram shown in Fig.~\ref{fig:gg-diagram}, which fills up the detector with some 19~TeV of energy per bunch train at 3~TeV, so a good selection is needed to be able to remove this background from the recorded data.
At 3~TeV, the expected rate of this type of background is 3.2 $\gamma \gamma \to$ hadrons events per bunch crossing with invariant mass $M_{\gamma\gamma}>$2 GeV.

\subsection{ The luminosity spectrum}

\begin{figure}[h!]
\centering
\includegraphics[width=0.5\textwidth]{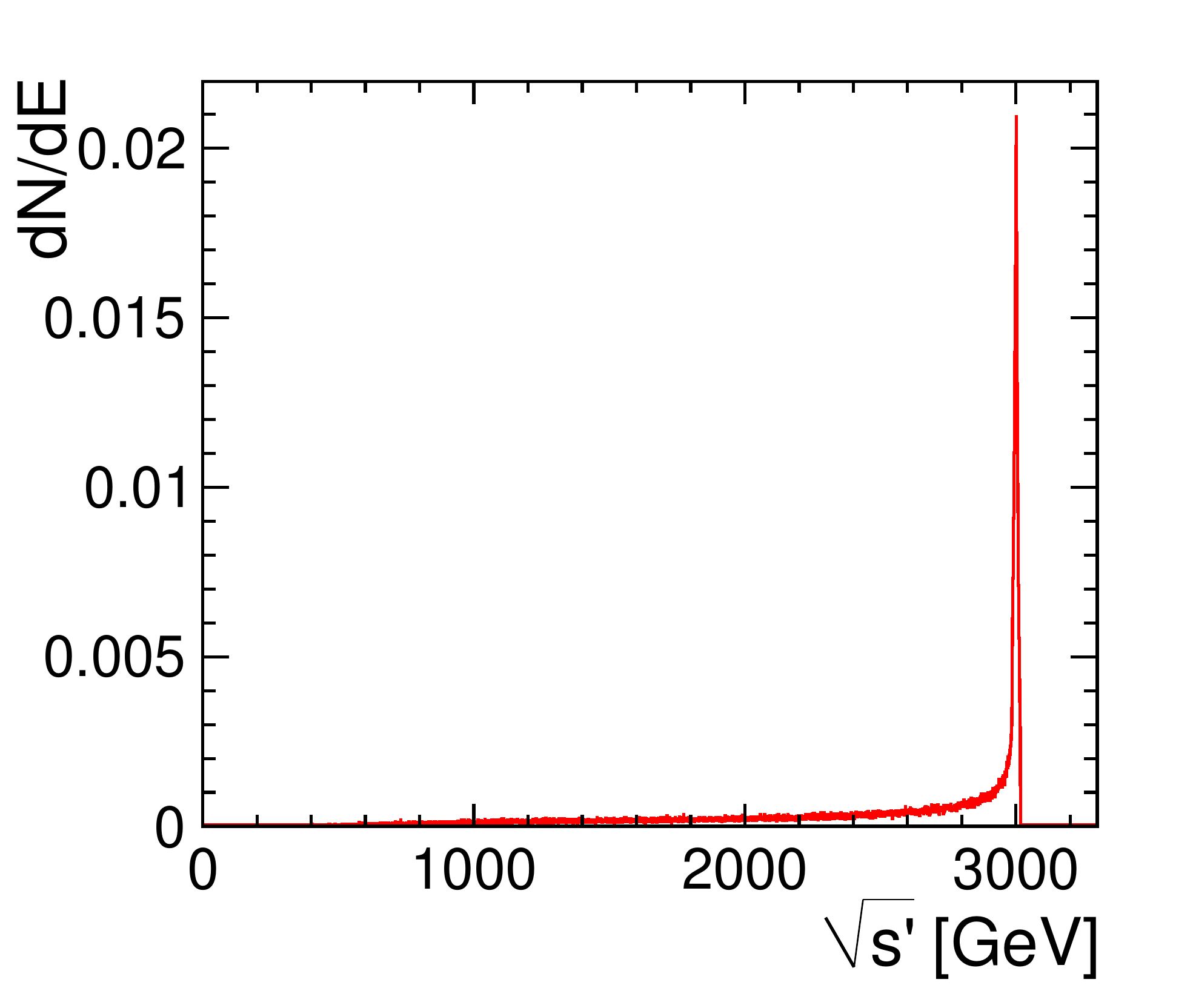}
\caption{The luminosity spectrum as a function of the effective cms,  $\sqrt{s^\prime}$, for a nominal CLIC cms energy $\sqrt{s}$ of 3~TeV.}
\label{fig:luminosity}
\end{figure}

The beamstrahlung and initial state radiation imply that not all the $e^+e^-$ collisions at CLIC will take place at the nominal cms energy, as the beam particles may radiate photons before their collision. Therefore, the effective cms energy of the collision, $\sqrt{s^\prime}$, will be less than the nominal cms energy of the machine, $\sqrt{s}$. Thus the luminosity spectrum will have a peak at $\sqrt{s}$ corresponding to collisions with no radiation prior to the $e^+e^-$ interaction and a long tail towards lower energies. The expected luminosity spectrum for the 3~TeV CLIC stage is shown in Fig.~\ref{fig:luminosity}.
At this energy, about 1/3 of the luminosity is in the most energetic 1\% fraction of the spectrum, and increases to above 1/2 at 500~GeV  ($\cal{L}_{\rm 0.01}$ in Table~\ref{tab:CLIC-stages}).

\section {CLIC detector concept}

\subsection{Impact of CLIC conditions on the detector}

The presence of beam-induced backgrounds impacts on the detector design. The $e^+e^-$ pairs from incoherent pair production give rise to high occupancies in certain regions of the detector, in particular the forward calorimetry, the vertex detector and the inner tracking regions. In the vertex and tracking detectors, high occupancies can be mitigated by appropriate shielding for back-scattered particles from the very forward regions and by the use of very small cell sizes, sometimes even smaller than strictly needed for detector resolution purposes. Background particles from $\gamma\gamma \to$ hadron events lead to significant energy depositions in the calorimeter systems, at the level of 20 TeV per bunch train at 3 TeV centre-of-mass energy. A scheme has been developed to suppress these energy depositions in the data. The scheme relies on precise hit timing, at the level of 1ns in the calorimetry and 10 ns in the tracking, combined with fine-grained calorimetry. This allows to reconstruct all particles, whether from background or from the physics event under study, using particle flow analysis. In this way the time of creation of each particle can be calculated very accurately, allowing for the rejection of out-of-time particles that do not match the time of the physics event.

\begin{figure}[h!]
\centering
\includegraphics[width=0.8\textwidth]{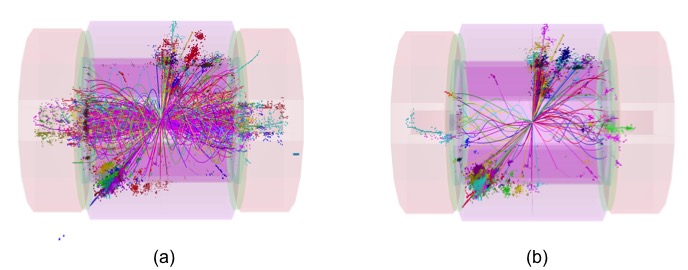}
\caption{An 8-jet event overlayed with background (a) and after filtering (b).}
\label{fig:before-after}
\end{figure}

The effectiveness of the above procedure is illustrated in Fig.~\ref{fig:before-after}, showing an event from the reaction 
$e^+e^-\to H^+H^-\to t\bar{b}b\bar{t}\to {\rm 8 \ jets}$  at 3 TeV before and after beam-induced background suppression according to this procedure.

\subsection {Detector needs from physics aims}

\subsubsection{Momentum resolution}

One of the physics subjects to be discussed in Section 5 is Higsstrahlung, $e^+e^- \to H Z$. This process, in which the Higgs is reconstructed as the particle recoiling off the $Z$ in the final state, is unique in allowing a model-independent determination of Higgs mass and coupling without any assumptions about the invisible decay of the Higgs. The $Z$ would be reconstructed from its muon decay and thus one needs a very good momentum resolution. Another physics process needing high momentum resolution would be the search for a smuon through the muon momentum endpoint. These two and other similar studies require a momentum resolution of 
\begin{equation}
\sigma_{p_T}/p_T^2 \sim 2 \times 10^{-5} \ {\rm GeV}^{-1} .
\label{eq:mom-res}
\end{equation}

\subsubsection{Jet-energy resolution}

In order to reconstruct $W$ and $Z$ from di-jets and be able to separate them, it is important to have excellent jet-energy resolution. For instance, for a 2.5\% resolution on the mass, a 3.5\% resolution on the jet energy is needed. Therefore the requirement for high energy jets would be
\begin{equation}
\sigma_E/E \leq {\rm 3.5}\% 
\label{eq:je-res}
\end{equation}
for jets with $E>$100~GeV ($\leq {\rm 5}$\% at 50~GeV).

\subsubsection{Impact parameter resolution}

In order to measure the Higgs branching ratios, it is important to be able to tag jets from $c$ and $b$ quarks. This calls for a good ability to reconstruct secondary vertices. This leads to the following requirement on the impact parameter resolution:
\begin{equation}
\sigma_{r\phi} = 5 \oplus 15 /(p[{\rm GeV}] sin^{\frac{3}{2}}\theta) \ \mu{\rm m} .
\label{eq:ip-res}
\end{equation}

\subsubsection{Angular coverage}

For many of the physics studies, the ability to tag forward-going electrons down to very small polar angle is important. This is discussed in the forward calorimetry section.

\newpage
\subsection{A possible CLIC detector}

\begin{figure}[h!]
\centering
\includegraphics[width=0.8\textwidth]{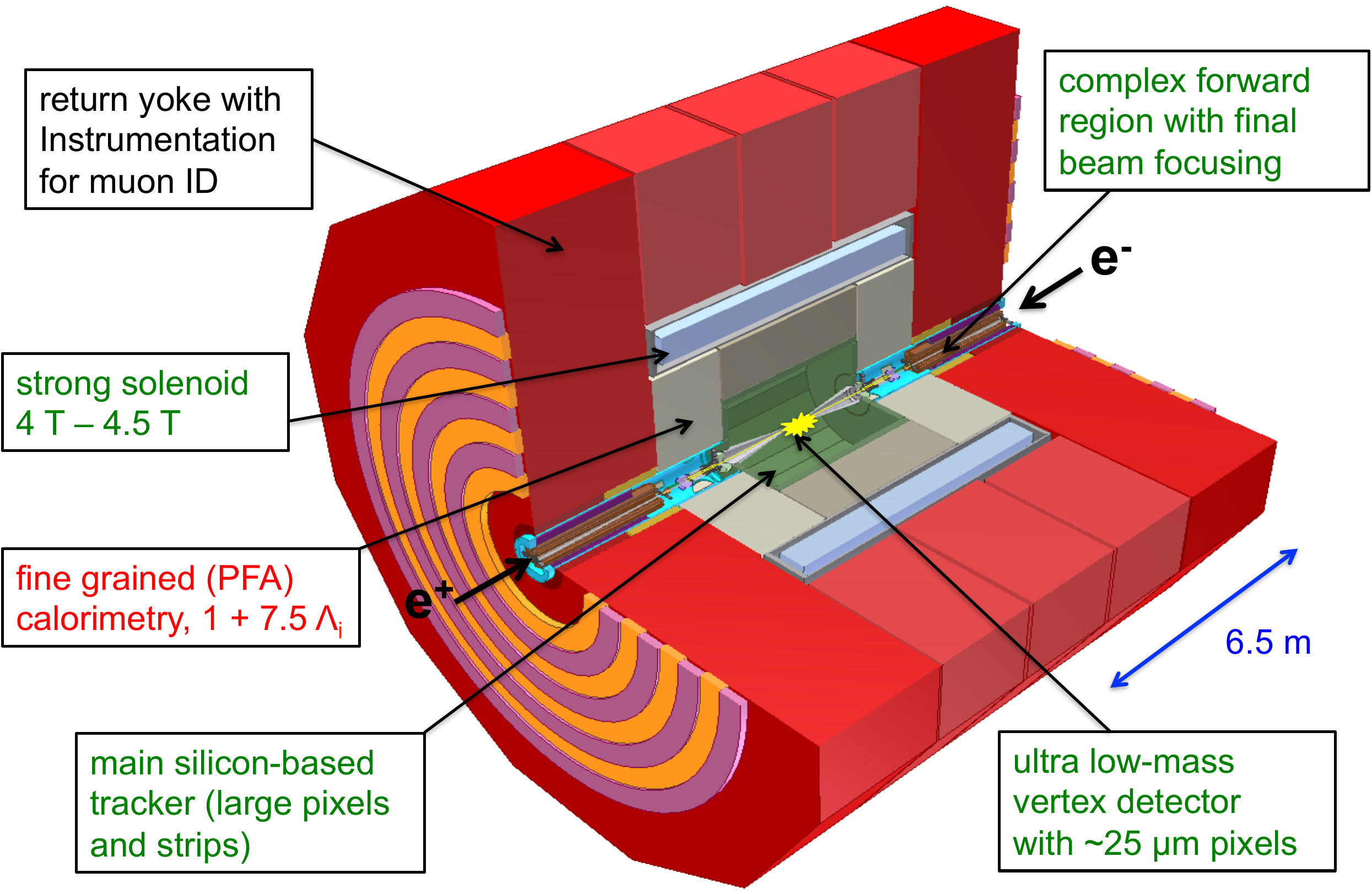}
\caption{A possible CLIC detector, derived from the two ILC detector concepts. The text in the framed boxes gives some explanation on the different components of the detector.}
\label{fig:CLICdp-det}
\end{figure}

The CLIC detector concept is derived from the ILC detector concepts, ILD~\cite{ILD} and SiD~\cite{SiD}, and is adapted to the conditions at CLIC. A schematic view of the CLIC detector is shown in Fig.~\ref{fig:CLICdp-det}. It is composed of barrel and endcap sections. Going outwards from the interaction point, the detector comprises: a low-mass vertex detector, a silicon-based main tracker, a fine-grained electromagnetic calorimeter (ECAL), a fine-grained hadron calorimeter (HCAL), a large 4 to 5 Tesla superconducting solenoid and an instrumented return yoke with muon identification layers. In the forward direction, near the incoming and outgoing beams, compact fine-grained forward calorimeters, LumiCal and BeamCal, are located. The final focusing quadrupoles for the incoming beams are also placed inside the detector volume. The overall length of the detector is approximately 13 m for a height of 14 m. 
In the following, a few points on some important items of the CLIC detector.

\subsubsection{Vertex detector}

The vertex detector is an essential part of the tracking system. Moreover, the vertex detector provides essential information on flavour tagging through the measurement of displaced vertices. The vertex detector comprises three double-layers of pixel detectors in the barrel region as well as in the forward disc region. A single-point resolution for each individual layer of 3~$\mu$m is required, and time-stamping for individual hits at the level of 10~ns. Occupancies from beam-induced backgrounds will be high and will limit the maximum allowed pixel surface. Overall a vertex detector of 2~Gpixels for an individual pixel surface of 25$\times$25~$\mu$m$^2$ is foreseen. The maximum amount of material amounts to 0.2\%$X_0$ per layer. The radiation levels will be significantly less (factor 10$^{-4}$) than at LHC.

\begin{figure}[h!]
\centering
\includegraphics[width=0.5\textwidth]{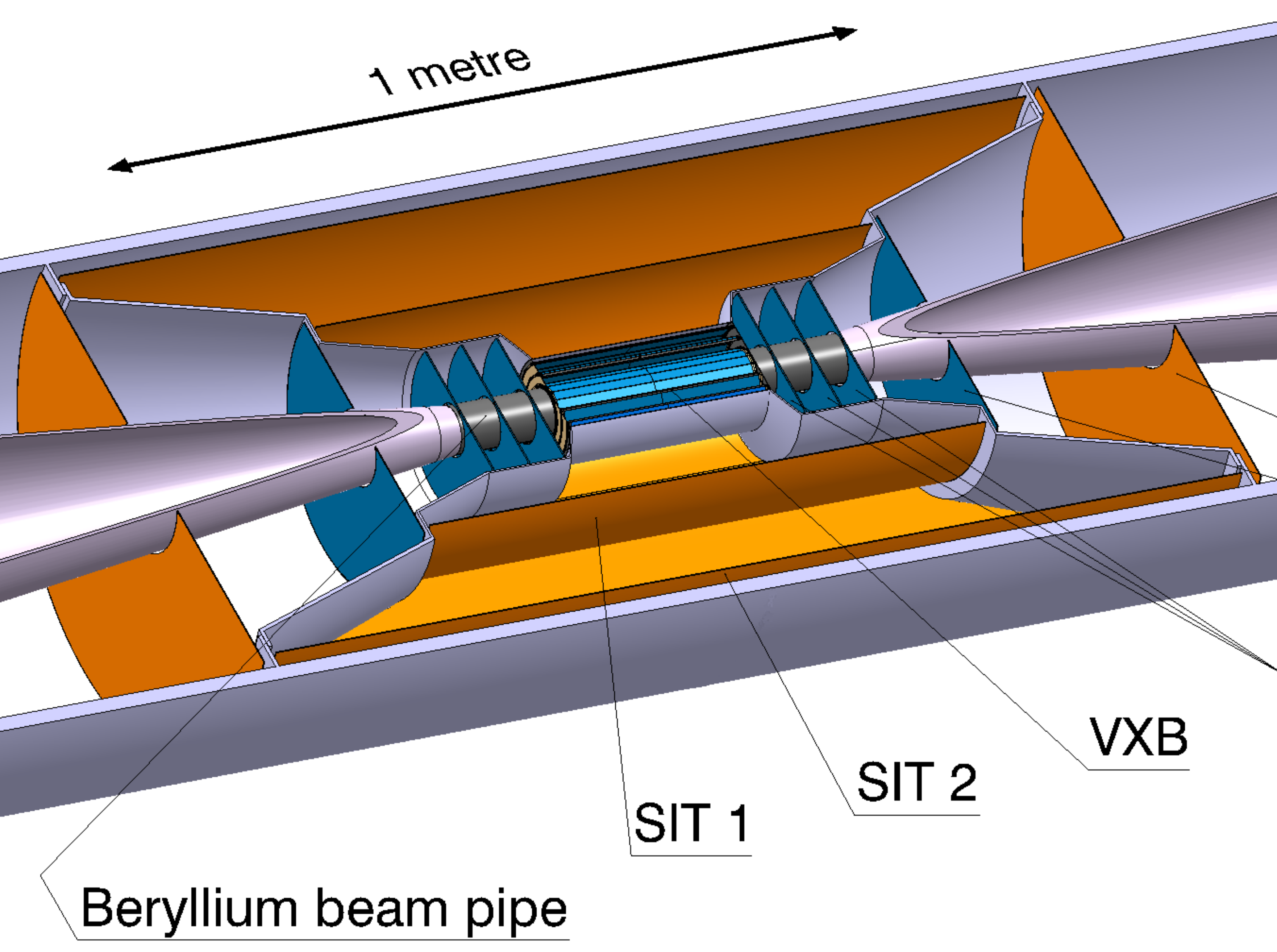}
\caption{A possible vertex detector layout.The two inner silicon barrel strips layers are SIT1 and SIT2, and the vertex barrel is VXB.}
\label{fig:VTX-det}
\end{figure}

To meet these requirements, very thin hybrid detectors composed of ~50~$\mu$m thin resistive silicon sensors and 50~$\mu$m ASICs are under development. To further meet the low-mass requirement, power pulsing at the 50~Hz beam train rate is foreseen. This will keep the heat dissipation at the level of 50 mW/cm$^2$ and will allow for air cooling. A vertex detector having these qualities could look like the one shown in Fig.~\ref{fig:VTX-det}.
The CLIC vertex detector is the subject of a very active detector R\&D project~\cite{VTX}.

\subsubsection{Tracker}

The main tracker at CLIC will be based on silicon technology. It will comprise large pixel elements as well as silicon microstrips, adapted to local occupancy levels.  Single point resolutions near 7~$\mu$m are required as well as time-stamping for individual hits at the level of 10~ns. The design of a silicon tracker concept and the assessment of the corresponding hardware implementation options for CLIC are still under study.

\subsubsection{Calorimetry and PFA}

\begin{figure}[h!]
\centering
\includegraphics[width=0.5\textwidth]{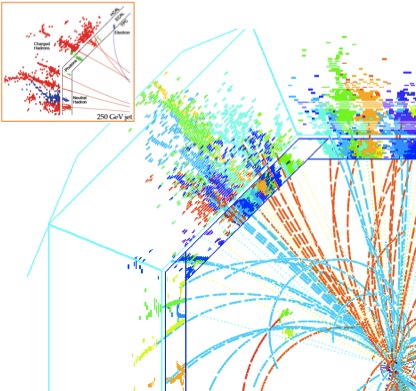}
\caption{Simulation of a 250~GeV jet produced in the ILD detector, consisting of different types of particles reconstructed using particle-flow analysis. The insert at the top left shows the separation of the types of particles.}
\label{fig:PFA}
\end{figure}
The jet energy resolution and the beam-induced background rejection drive the overall detector design. As already mentioned above, the good jet energy  resolution and background rejection are achieved through fine-grained calorimetry and by a particle-flow analysis (PFA)~\cite{PFA}.  Figure~\ref{fig:PFA} is an illustration of how one recognises individual particles in jets using the PFA algorithm. Typically 60\% of a jet is composed of charged particles, 30\% of photons and 10\% of long-lived neutral hadrons. Charged particles profit from the excellent momentum resolution in the tracker, while photons are well measured in the ECAL. Neutrons rely on measurements in the HCAL, which is less good due to hadronic shower fluctuations. By reconstructing individual particles in jets, the best information can be used for each of them and a good overall jet energy resolution can be obtained. The combination of hardware and software is providing the wanted result.

\subsection{CLIC forward calorimetry}

\begin{figure}[h!]
\centering
\includegraphics[width=0.8\textwidth]{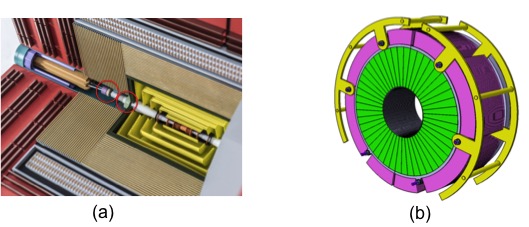}
\caption{ (a) Part of a CLIC detector showing the two forward calorimeters, LumiCal and BeamCal inside the red circles. (b) LumiCal, a compact luminosity calorimeter, composed of Tungsten layers and silicon sensors.}
\label{fig:FCAL}
\end{figure}

Two small compact calorimeters are planned in the forward region (Fig.~\ref{fig:FCAL}, the two red circles), with the aim to measure the luminosity, to provide electron veto capabilities at small angles as well as a fast beam monitoring system and to complement the hermeticity of the full detector. 
The luminosity detector, LumiCal, covering the angular region of 38-110~mrad,  will perform precision measurements~\cite{FCAL-JINST} of the Bhabha event rate and thereby measure the luminosity.
The very forward coverage of BeamCal, 10-40~mrad, enhances the hermiticity of the detector.  The  BeamCal is subject to high doses of radiation, even though by four orders of magnitude less than those in the forward region of LHC. It requires the BeamCal to use radiation-hard sensors; GaAs sensors are presently proposed.

\newpage
\section{CLIC physics capabilities}

In this section we will present examples of benchmark studies~\cite{bench-studies}  performed with 
GEANT4-based~\cite{GEANT4} full detector simulations, with overlay of $\gamma\gamma$ background. The analyses include SM physics backgrounds. So far, most of the studies concentrated on Higgs physics, while some were dedicated to top physics and new physics with CLIC.

\subsection{Higgs physics at CLIC}

\begin{figure}[h!]
\centering
\includegraphics[width=0.6\textwidth]{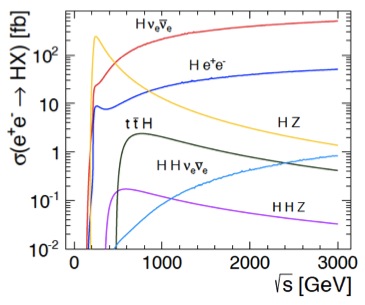}
\caption{ Cross sections for Higgs production processes as a function of the cms energy, $\sqrt{s}$. The different processes are indicated in the figure. The values shown are for unpolarised beams and do not include effects of initial-state radiation or beamstrahlung.}
\label{fig:higgs-xsect}
\end{figure}

The dominant cross section for Higgs production at low energies is the Higgsstrahlung process,
\begin{equation}
e^+ + e^- \to Z + H \ ,
\label{eq:higgsstrahlung}
\end{equation}
which peaks at $ \sqrt{s} \sim$240~GeV, and then falls off with cms energy. The $W(Z)$ fusion process,
\begin{equation}
e^+ + e^- \to H + \nu_e + \bar{\nu}_e \ ,
\label{eq:Wfusion}
\end{equation}
increases with $\sqrt{s}$ with the $WW$ fusion process having about 10 times higher cross section than the $ZZ$ fusion ($He^+e^-$). The Higgs production cross sections as a function of the cms energy are presented in Fig.~\ref{fig:higgs-xsect}, and a table of the expected number of produced events of the three dominant Higgs processes, for the integrated luminosities at the different cms energies (usually over a period of four years), are listed in Table~\ref{tab:events}. The expected number of events in the table take into account that the beam has an energy spectrum and not all interactions happen at the nominal cms energy.
\vspace{-0.5cm}
\begin{table}[h!]
\centering
\caption{Expected number of produced events of the three leading Higgs processes for three cms energies and for their respective integrated luminosities. The effects of initial-state radiation and the beam spectrum are taken into account. Unpolarised beams are assumed}
\label{tab:events}      
\begin{tabular}{|l|c|c|c|}
\hline
 $\sqrt{s}$ (TeV) & 0.35 & 1.4 & 3 \\ \hline \hline
Int. Lumi. (fb$^{-1}$) & 500 & 1,500 & 2,000 \\ \hline
\# of $ZH$ evts &68,000&20,000&11,000 \\\hline
\# of $H \nu_e \bar{\nu}_e$ evts &17,000 & 370,000 & 830,000 \\\hline
\# of $H e^+ e^-$ evts & 3,700 & 37,000& 84,000 \\\hline
\end{tabular}
\end{table}

\begin{figure}[h!]
\centering
\includegraphics[width=0.8\textwidth]{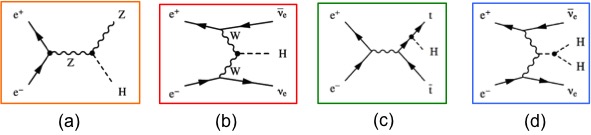}
\caption{ Diagrams of four Higgs-production processes. (a) Higgsstrahlung, (b) $WW$ fusion, (c) radiation off top-quarks, and (d) double-Higgs production.}
\label{fig:higgs-diagrams}
\end{figure}

Diagrams of four processes, Higgstrahlung, $WW$ fusion, radiation off top-quarks and double-Higgs production, are presented in Fig.~\ref{fig:higgs-diagrams}.

\subsubsection{Higgsstrahlung}

\begin{figure}[h!] 
\setlength{\unitlength}{1cm}
\centering
\begin{picture}(8,5.4)
\put(0,-0.35){\includegraphics[width=0.5\textwidth]{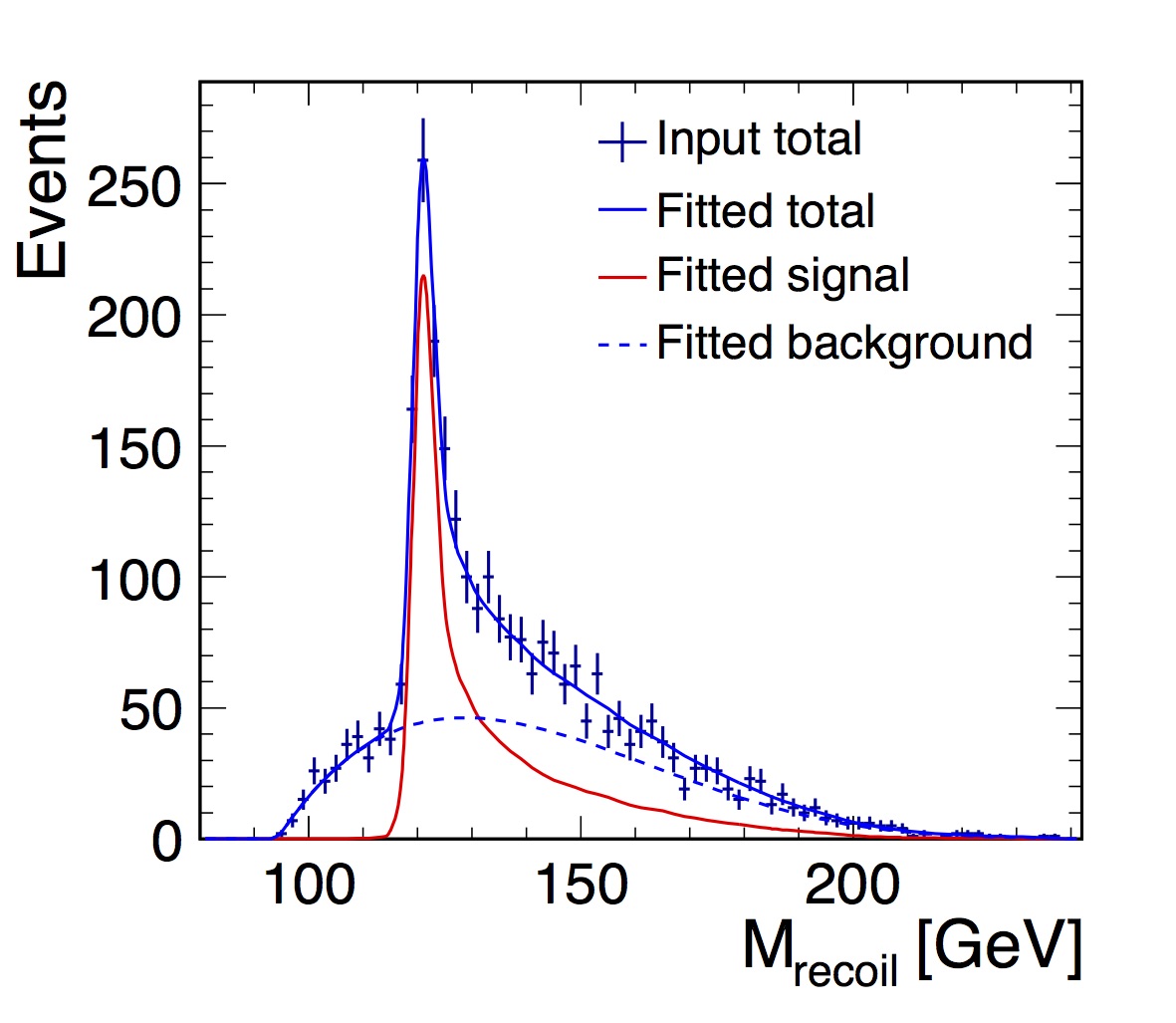}}
\put(4,2.8){ {\bf CLICdp}}
\end{picture}
\caption{ Simulation of the recoiling mass distribution to a $Z$ in Higgsstrahlung at a cms energy of 350~GeV using an integrated luminosity of 500~fb$^{-1}$. The $Z$ is reconstructed through its $\mu\mu$ decay mode.}
\label{fig:MH-higgsstrahlung}
\end{figure}

The Higgsstrahlung reaction, (\ref{eq:higgsstrahlung}) and Fig.~\ref{fig:higgs-diagrams}(a), is a unique process for $e^+e^-$ interactions in the sense that it provides a model-independent measurement of the Higgs mass and production cross section, without the need to assume anything about the decay branching ratios of the Higgs. The Higgs properties are determined in a missing-mass type of experiment by measuring it as a recoil mass to the $Z$.
The $Z$ can be best reconstructed through the $\mu\mu$ decay mode by using the tracker. A simulation of the recoil-mass distribution from the reconstructed $Z \to \mu\mu$ is shown in Fig.~\ref{fig:MH-higgsstrahlung}, demonstrating the ability of a precision measurement of the Higgs mass and width.

Such a measurement determines the absolute value of the $HZZ$ coupling, $g_{HZZ}$. Using the $Z \to \mu\mu$ and $Z \to ee$ decay modes, together with the hadronic $Z \to qq$ modes, which has a much higher decay rate and thus adds much to the statistical accuracy, one can determine~\cite{TM} the $g_{HZZ}$ coupling with an accuracy of
\begin{equation}
\Delta g_{HZZ} = \pm0.8\%.
\end{equation}

\subsubsection{$WW$ fusion}

The $WW$ fusion reaction, which has the highest cross section for Higgs production above $\sqrt{s}$ of 350~GeV (Fig.~\ref{fig:higgs-xsect}), is depicted in Fig.~\ref{fig:higgs-diagrams}(b). This process will give precise cross section measurements in the different decay modes $\tau\tau$, $\mu\mu$, $q q$, etc.

\subsubsection{Radiation off top-quarks}

The process $e^+ e^- \to t \bar{t} H$, presented in the diagram of Fig.~\ref{fig:higgs-diagrams}(c), is preferably measured at  energies above 700~GeV. It is a very important process for the measurement of the Yukawa coupling. By measuring the Higgs through its $b\bar{b}$ decay, the Yukawa coupling is expected~\cite{bench-studies} to be measured with an uncertainty of 4.5\% at 1.4~TeV.

\subsubsection{Double Higgs production}

\begin{figure}[h!]
\centering
\includegraphics[width=0.3\textwidth]{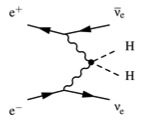}
\caption{ Diagram for double-Higgs production, sensitive to the quartic coupling $g_{HHWW}$.}
\label{fig:quartic}
\end{figure}

The double-Higgs process, $e^+ e^- \to H H \nu_e \bar{\nu}_e$, presented by the diageam in Fig.~\ref{fig:higgs-diagrams}(d), is sensitive to the Higgs self coupling, $\lambda$. However, the same final state can also be produced  via the diagram shown in Fig.~\ref{fig:quartic}. Therefore the measurement is also sensitive to the Higgs quartic coupling, $g_{HHWW}$.
The double Higgs cross section is $\sigma(HH\nu_e\bar{\nu}_e) $ = 0.15 (0.59)~fb at 1.4 (3)~TeV, so it is crucial to have high energy and high luminosity for this measurement.
\begin{table}[h!]
\centering
\caption{Uncertainties on the measurements of the self and quartic Higgs couplings for two different cms energies.}
\label{tab:couplings}      
\begin{tabular}{|l|c|c|}
\hline
 &  1.4~TeV & 3~TeV \\ \hline \hline
$\Delta(g_{HHWW}) $& 7\% (prel.) & 3\% (prel.) \\ \hline
$\Delta(\lambda) $&32\%&16\% \\\hline
$\Delta(\lambda)$, P(e$^-$)=80\% & 24\% & 12\% \\\hline
\end{tabular}
\end{table}
 The expected precision with which the self and quartic Higgs coupling can be determined at 1.4 and 3~TeV are given in Table~\ref{tab:couplings}. Having an electron beam with 80\% polarisation can substancially improve the precision of the measurement.

\subsubsection{Summary of Higgs measurements and global fits}

A summary of the projected statistical accuracies for Higgs measurements at three different CLIC cms energies is given in Table~\ref{tab:summary}. Various Higgs production and decay modes are listed. These are individually accessible thanks to the excellent tracking, jet energy measurement and flavour tagging capabilities of the CLIC detector. Neglecting the Higgs decay into light quarks, the gg decay mode of the Higgs is estimated. 
\begin{table}[h!]
\centering
\caption{Statistical accuracy on the measurements of Higgs-related observables from the processes of Higgsstrahlung, $WW$ fusion,  radiation off top quarks and double-Higgs production for three different cms energies with integrated luminosities as listed.}
\label{tab:summary}      
\begin{tabular}{|l|c|l|c|c|c|}
\hline
 Channel&  Measurement & Observable & 350~GeV& 1.4~TeV& 3.0~TeV \\ 
 & & & 500~fb$^{-1}$ & 1.5~ab$^{-1}$ & 2.0~ab$^{-1}$ \\ \hline \hline
$ZH$ & Recoil mass distribution & $m_H$ & 120~MeV & - & - \\ \hline
$ZH$ & $\sigma(HZ)\times BR(H\to {\rm invisible})$ & $\Gamma_{inv}$ & 0.6\% & - & - \\ \hline
$ZH$ & $H \to b\bar{b}$ mass distribution & $m_H$ & tbd & - & - \\ \hline
$H\nu_e\bar{\nu}_e$ & $H \to b\bar{b}$ mass distribution & $m_H$ & - & 40~MeV& 33~MeV \\ \hline \hline
$ZH$ & $\sigma(ZH)\times BR(Z\to l^+l^-)$ & $g^2_{HZZ}$ & 4.2\% & - & - \\ \hline
$ZH$ & $\sigma(ZH)\times BR(Z\to q\bar{q})$ & $g^2_{HZZ}$ & 1.8\% & - & - \\ \hline
$ZH$ & $\sigma(ZH)\times BR(H\to b\bar{b})$ & $g^2_{HZZ}g^2_{Hbb}/\Gamma_H$ & 1\% & - & - \\ \hline
$ZH$ & $\sigma(ZH)\times BR(H\to c\bar{c})$ & $g^2_{HZZ}g^2_{Hcc}/\Gamma_H$ & 5\% & - & - \\ \hline
$ZH$ & $\sigma(ZH)\times BR(H\to gg)$ &  & 6\% & - & - \\ \hline
$ZH$ & $\sigma(ZH)\times BR(H\to \tau^+\tau^-)$ & $g^2_{HZZ}g^2_{H\tau\tau}/\Gamma_H$ & 5.7\% & - & - \\ \hline
$ZH$ & $\sigma(ZH)\times BR(H\to WW^*)$ & $g^2_{HZZ}g^2_{HWW}/\Gamma_H$ & 2\% & - & - \\ \hline
$ZH$ & $\sigma(ZH)\times BR(H\to ZZ^*)$ & $g^2_{HZZ}g^2_{HZZ}/\Gamma_H$ & tbd & - & - \\ \hline
$H\nu_e\bar{\nu}_e$ & $\sigma(H\nu_e\bar{\nu}_e)\times BR(H\to b\bar{b})$ & $g^2_{HWW}g^2_{Hbb}/\Gamma_H$ & 3\% & 0.3\% & 0.2\% \\ \hline
$H\nu_e\bar{\nu}_e$ & $\sigma(H\nu_e\bar{\nu}_e)\times BR(H\to c\bar{c})$ & $g^2_{HWW}g^2_{Hcc}/\Gamma_H$ & -  & 2.9\% & 2.7\% \\ \hline
$H\nu_e\bar{\nu}_e$ & $\sigma(H\nu_e\bar{\nu}_e)\times BR(H\to gg)$ &  & - & 1.8\% & 1.8\% \\ \hline
$H\nu_e\bar{\nu}_e$ & $\sigma(H\nu_e\bar{\nu}_e)\times BR(H\to \tau^+\tau^-)$ & $g^2_{HWW}g^2_{H\tau\tau}/\Gamma_H$ & - & 3.7\% & tbd \\ \hline
$H\nu_e\bar{\nu}_e$ & $\sigma(H\nu_e\bar{\nu}_e)\times BR(H\to \mu^+\mu^-)$ & $g^2_{HWW}g^2_{H\mu\mu}/\Gamma_H$ & - & 38\% & 16\% \\ \hline
$H\nu_e\bar{\nu}_e$ & $\sigma(H\nu_e\bar{\nu}_e)\times BR(H\to \gamma\gamma)$ &  & - & 15\% & tbd \\ \hline
$H\nu_e\bar{\nu}_e$ & $\sigma(H\nu_e\bar{\nu}_e)\times BR(H\to Z\gamma)$ &  & - & 42\% & tbd \\ \hline
$H\nu_e\bar{\nu}_e$ & $\sigma(H\nu_e\bar{\nu}_e)\times BR(H\to WW^*)$ & $g^4_{HWW}/\Gamma_H$ & tbd & 1.1\% & 0.8\% \\ \hline
$H\nu_e\bar{\nu}_e$ & $\sigma(H\nu_e\bar{\nu}_e)\times BR(H\to ZZ^*)$ & $g^2_{HWW}g^2_{HZZ}/\Gamma_H$ & - & 3\% & 2\% \\ \hline
$He^+e^-$ & $\sigma(He^+e^-)\times BR(H\to b\bar{b} )$ & $g^2_{HZZ}g^2_{Hbb}/\Gamma_H$ & - & 1\%& 0.7\% \\ \hline \hline
$t\bar{t}H$ & $\sigma(t\bar{t}H)\times BR(H\to b\bar{b} )$ & $g^2_{HZZ}g^2_{Hbb}/\Gamma_H$ & - & 8\%& tbd \\ \hline 
$HH\nu_e\bar{\nu}_e$ & $\sigma(HH\nu_e\bar{\nu}_e)$ &$g_{HHWW}$ & - & 7\%& 3\% \\ \hline 
$HH\nu_e\bar{\nu}_e$ & $\sigma(HH\nu_e\bar{\nu}_e)$ &$\lambda$ & - & 32\%& 16\% \\ \hline 
$HH\nu_e\bar{\nu}_e$ & with -80\% $e^-$ polarisation &$\lambda$ & - & 24\%& 12\% \\ \hline 
\end{tabular}
\end{table}
The first part of the table gives the precision with which the mass and width of the Higgs can be determined from Higgsstrahlung at 350~GeV and from the $b\bar{b}$  decay mode of the Higgs produced in the high statistics $WW$  fusion process at the higher energies. The second part gives the precision of the cross section measurements for the different decay modes of the two processes. The third part gives similar information from the process of radiation off the top quark and the double-Higgs reaction. For all cases, except the last row, unpolarised beams are assumed. For
the Higgs self-coupling observable $\lambda$, the projected accuracy for a 80\% polarised electron beam is also presented.

For the branching ratios, the measurement precision refers to the expected statistical uncertainty on the product of the relevant cross section and branching ratio; this is equivalent to the expected statistical uncertainty of the product of couplings divided by the Higgs width $\Gamma_H$ as listed in the third column. Most measurements listed in the table are based on full detector simulations, including overlay of beam-induced background from $\gamma\gamma$ to hadrons as well as background from physics events. See~\cite{bench-studies} for more detailed information.

As mentioned above, the Higgsstrahlung process provides model-independent absolute values for the $g_{HZZ}$ coupling. By using its value together with the ratios of observables obtained in all other measurements, one can obtain all other couplings in a model-independent way. This is done in a global fit, using the accuracies of the measurements quoted in Table~\ref{tab:summary}. First only the 350~GeV measurements are used, then those of 
1.4~TeV are added, and finally also the 3~TeV results are included, increasing the statistical accuracy of the results. In this global fit, one assumes an 80\% polarised electron beam above 1 TeV. Results are presented in Table~\ref{tab:global}.
\begin{table}[h!]
\centering
\caption{The precision of the Higgs mass and Higgs-related couplings obtained from a global fit to the measurements presented in Table~\ref{tab:summary}.}
\label{tab:global}      
\begin{tabular}{|l|c|c|c|}
\hline
 Parameter & 350~GeV &  1.4~TeV & 3~TeV \\ 
 &  500~fb$^{-1}$ & 1.5~ab$^{-1}$ & 2.0~ab$^{-1}$ \\ \hline \hline
$m_H [MeV]$ & 120 & 30 & 20 \\ \hline
$\lambda$[\%] & - & 24 & 11\\ \hline  \hline
$\Gamma_H$[\%] & 5.0 & 3.6 & 3.4 \\ \hline
$g_{HZZ}$[\%] & 0.8 & 0.8 & 0.8 \\ \hline
$g_{HWW}$[\%] & 1.8 & 0.9 & 0.9 \\ \hline
$g_{Hbb}$[\%] & 2.0 & 1.0 & 0.9 \\ \hline
$g_{Hcc}$[\%] & 3.3 & 1.4 & 1.1 \\ \hline
$g_{Htt}$[\%] & - & 4.1 & 4.1 \\ \hline
$g_{H\tau\tau}$[\%] & 3.5 & 1.6 & $<$1.5 \\ \hline
$g_{H\mu\mu}$[\%] & - & 14 & 5.6 \\ \hline
$g_{Hgg}$[\%] & 3.6 & 1.1 & 1.0 \\ \hline
$g_{H\gamma\gamma}$[\%]& - & 5.7 & $<$5.7 \\ \hline
\end{tabular}
\end{table}

The results obtained in the global fit can now be used in constrained LHC-style fits to obtain the precision on the Higgs coupling scale factors which indicate the deviation from the SM. The LHC-style fits are model dependent in the sense that one assumes that there are no invisible decay modes of the Higgs. The coupling scale factors are defined as
\begin{equation}
\kappa^2_i = \frac{\Gamma_i}{\Gamma^{SM}_i} ,
\end{equation}
where $\Gamma_i$ and $\Gamma^{SM}_i$, respectively,  are the measured and the SM prediction for the partial width of a visible Higgs decay mode $i$. The total Higgs decay width, $\Gamma_{H, model}$ is the sum of the visible partial decay widths,
\begin{equation}
\Gamma_{H,model} = \Sigma_i{ \kappa^2_i BR_i}
\end{equation}
Fits of this type yield sub-\% precision for most couplings~\cite{bench-studies}.

\subsection{Top physics at CLIC}

The top pair is produced in the reaction
\begin{equation}
e^+ + e^- \to t + \bar{t} \to (W^+ + b) + (W^- + \bar{b}) .
\end{equation} 
The existing studies concentrate on the top mass measurements~\cite{Seidel}. The coupling to the Higgs are part of the Higgs studies presented above.

There are two ways to measure the top mass at CLIC. One is the classical way of reconstructing the $t\bar{t}$ pair through their all-hadronic decay mode. The second is by performing a careful energy scan and measuring the top quark pair production cross section in the vicinity of the $t\bar{t}$ threshold. The cross section is very sensitive to the value of the top mass. Such a scan also yields information about $\alpha_S$.

\begin{figure}[h!]
\centering
\includegraphics[width=1.0\textwidth]{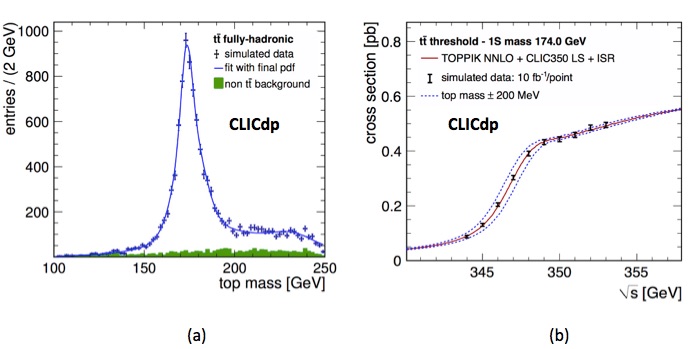}
%\includegraphics[width=0.46\textwidth]{figs/TopMassFHNoRes.pdf}
%\includegraphics[width=0.46\textwidth]{figs/TopThreshold.pdf}
%\vspace*{-2cm}
\caption{(a) Reconstructed top quark mass in the fully-hadronic decay mode for an integrated luminosity of 100~fb$^{-1}$ at 500~GeV. The solid line is the result of a fit to the invariant mass distribution. (b) A scan of the top quark pair production threshold in 10 energy steps, each with a luminosity of 10~fb$^{-1}$. The lines are NNLO calculations of the cross section, using different top masses.}
\label{fig:ttbar-threshold}
\end{figure} 

Figure~\ref{fig:ttbar-threshold}(a) shows the reconstructed invariant mass distribution of fully-hadronic top-pair decays for an integrated luminosity of 100~fb$^{-1}$ at 500~GeV. A mass fit, which also takes into account the non-$t\bar{t}$ background, results in a statistical error of 80~MeV on the mass. 

The energy scan around the $t\bar{t}$ threshold is shown in Fig.~\ref{fig:ttbar-threshold}(b). The data points represent measurements at 10 energy steps, each with a luminosity of 10~fb$^{-1}$. The NNLO calculations of the cross section by Hoang and Teubner~\cite{HT}, including effects of the luminosity spectrum and initial state radiation (ISR), are sensitive to the top mass and to the value of $\alpha_S$ used in the calculation. Using a two-dimensional fit, one can determine the mass with a statistical uncertainty of 33~MeV, and the value of $\alpha_S$ with an uncertainty of $\pm$0.0009. Taking into account that the result has to include also systematic uncertainties like those coming from theory normalisation and beam-energy systematics, one can expect to determine the top mass with a total uncertainty of about 100~MeV.

\subsection{Physics Beyond the Standard Model (BSM)}

\subsubsection{Extension of the Higgs sector}

Supersymmetry, a proposed extension of the SM, may explain several open questions in particle physics and cosmology. The Minimal Supersymmetric Standard Model (MSSM) is one of the best studied candidates for physics BSM. 
At lowest order, where the MSSM Higgs sector is $\cal{CP}$-conserving, five physical Higgs bosons are required. They are the light and heavy $\cal{CP}$-even Higgs bosons, $h$ and $H$, the $\cal{CP}$-odd Higgs boson, $A$, and the charged Higgs boson pair, $H^\pm$.

\begin{figure}[h!]
\centering
\includegraphics[width=1.0\textwidth]{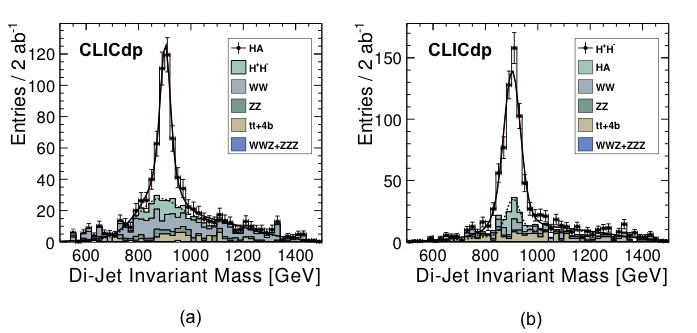}
\caption{Reconstructed MSSM Higgs mass in the process (a) $e^+e^- \to HA$ and (b) $e^+e^- \to H^+H^-$. The corresponding background channels are shown as well.}
\label{fig:MSSM-Higgs}
\end{figure}

The neutral and charged MSSM heavy Higgs bosons~\cite{MSSMHiggs} are examples of MSSM benchmark studies for a 3~TeV CLIC with integrated luminosity of 2~ab$^{-1}$. The expected cross sections are of the order of a few~fb at this energy and the pair production proceeds through the reaction
\begin{equation}
e^+ + e^- \to H + A \ ,
\end{equation}
and 
\begin{equation}
e^+ + e^- \to H^+ + H^- .
\end{equation}
They are studied~\cite{LL-CDR} by reconstructing their complex final state from their decays into $b\bar{b}b\bar{b}$ and $t\bar{b}t\bar{b}$, respectively. The di-jet invasriant mass of the two doublets, the neutral and the charged, are presented in Fig~\ref{fig:MSSM-Higgs}. In this model the doublets are almost degenerate in mass and their mass can be well determined.

Further studies~\cite{LL-CDR,CLIC} of Supersymmetrtic (SUSY) benchmark processes producing sleptons, charginos, neutralinos, squarks and staus have been performed with full-detector simulation with background overlaid. All the results at 3~TeV (1.4~TeV), based on an integrated luminosity of 2~ab$^{-1}$ (1.5~ab$^{-1}$), show that if these states exist, a large part of the SUSY spectrum can be measured at a $<$1\% level.

\subsection{LHC and CLIC}

Before concluding, it is of interest to show a comparison performed by De Roek, Ellis and Gianotti~\cite{Roek} of the discovery reach of various theory models for different colliders. The results shown in Table~\ref{tab:compare} lead to a conclusion that CLIC is an ideal machine for performing precision studies, for stand-alone discoveries and for extending our search for physics  beyond the SM.
\begin{table}[h!]
\centering
\caption{Discovery reach of various theory models for different colliders. LHC at $\sqrt{s}$ = 14~TeV assumes 100~fb$^{-1}$ of integrated luminosity, while HL-LHC is with 1~ab$^{-1}$, and CLIC at 3~TeV with up to 2~ab$^{-1}$. TGC is short for Tripple Gauge Coupling, and $\mu$ contact scale is short for LL $\mu$ contact interaction scale $\Lambda$ with $g$ = 1. From~\cite{Roek}.}
\label{tab:compare}      
\begin{tabular}{|l|c|c|c|}
\hline
New particle & LHC(14 TeV) & HL-LHC & CLIC (3~TeV) \\ \hline \hline
squarks [TeV] & 2.5 & 3 & $\leq$1.5 \\ \hline
sleptons [TeV] & 0.3 & - & $\leq$1.5 \\ \hline
$Z^\prime$ (SM couplings) [TeV & 5 & 7 & 20 \\ \hline
2 extra dims $M_D$ [TeV] & 9 & 12 & 20-30 \\ \hline
TGC (95\%) ($\lambda_\gamma$ coupling) & 0.001 & 0.0006 & 0.0001 \\ \hline
$\mu$ constant scale [TeV] & 15 & - & 60 \\ \hline
Higgs composite scale [TeV] & 5-7 & 9-12 & 70 \\ \hline
\end{tabular}
\end{table}

%\newpage
\section{Summary}

CLIC is currently the only mature option for a multi-TeV $e^+e^-$ collider. Very active R\&D projects are ongoing for accelerator, physics and detector. The construction and operation of the machine will proceed in stages in order to offer high luminosities over a broad energy range for a maximum physics coverage. The exact choice of the stages will be driven by LHC physics results. At present, possible stages at 350~GeV, 1.4 and 3~TeV are considered. At 350~GeV, precision Higgs and top measurements will be performed. At the higher energies, improved precision of many observables will be carried out and access to rare Higgs decays will be possible.  At the high energies the collider will serve as a discovery machine for BSM physics.

\section*{Acknowledgments}

I would like to thank my colleagues in the CLICdp collaboration for their support. I am particularly indepted to Lucie Linssen for her help in preparing this talk and this writeup. I would like to acknowledge the very special atmosphere created at this meeting by  Sonia Kabana and her team of co-organisers.

\end{document}